\documentclass[conference]{IEEEtran}
\IEEEoverridecommandlockouts
\usepackage{cite}
\usepackage{amsmath,amssymb,amsfonts}
\usepackage{float}
\usepackage{booktabs, makecell, multirow}
\usepackage{algorithm}
\usepackage{algorithmic}
\usepackage{graphicx, subcaption}
\usepackage{textcomp}
\usepackage{xcolor}
\usepackage{epsfig}
\usepackage{epstopdf}
\usepackage{balance}
\usepackage{tabularx}
\usepackage{adjustbox}
\usepackage{mathtools}
\usepackage[printonlyused]{acronym}

\acrodef{uav}[UAV]{unmanned air vehicle}
\acrodef{gs}[GS]{ground station}
\acrodef{cir}[CIR]{channel impulse response}
\acrodef{pdp}[PDP]{power-delay profile}
\acrodef{rms}[RMS]{root-mean-square}
\acrodef{ple}[PLE]{path loss exponent}
\acrodef{rms-ds}[RMS-DS]{\ac{rms} delay spread}
\acrodef{pn}[PN]{pseudo--noise}
\acrodef{los}[LOS]{line--of--sight}
\acrodef{a2g}[A2G]{Air--to--Ground}
\acrodef{a2a}[A2A]{Air--to--Air}
\acrodef{sdr}[SDR]{software-defined radio}
\acrodef{sbc}[SBC]{single board computer}
\acrodef{ssd}[SSD]{solid state drive}
\acrodef{gpsdo}[GPSDO]{GPS disciplined oscillator}
\acrodef{mp}[M]{measurement point}
\acrodef{txd}[TX-D]{transmitter drone}
\acrodef{rxd}[RX-D]{receiver drone}
\acrodef{swap}[SWaP]{size weight and power}
\acrodef{mpc}[MPC]{multipath components}
\acrodef{dpc}[DPC]{direct path component}
\acrodef{fspl}[FSPL]{free--space path loss}
\acrodef{fe2r}[FE2R]{flat-earth two-ray}

\newcommand{\FGR}[1]{Fig.~\ref{#1}}
\newcommand{\FGRu}[1]{Figure~\ref{#1}}

\newcommand{\TAB}[1]{Table~\ref{#1}}
\newcommand{\EQ}[1]{(\ref{#1})}

\def\BibTeX{{\rm B\kern-.05em{\sc i\kern-.025em b}\kern-.08em
    T\kern-.1667em\lower.7ex\hbox{E}\kern-.125emX}}
\begin{document}

\title{Measurement-based Channel Characterization for A2A and A2G Wireless Drone Communication Systems}

\author{\IEEEauthorblockN{Ubeydullah Erdemir\IEEEauthorrefmark{1}\IEEEauthorrefmark{2}, Batuhan Kaplan\IEEEauthorrefmark{1}, İbrahim Hökelek\IEEEauthorrefmark{1}\IEEEauthorrefmark{2}, Ali Görçin\IEEEauthorrefmark{1}\IEEEauthorrefmark{3}, Hakan Ali Çırpan\IEEEauthorrefmark{2}}

\IEEEauthorblockA{\IEEEauthorrefmark{1}Communications and Signal Processing Research (HİSAR) Lab., T{\"{U}}B{\.{I}}TAK-B{\.{I}}LGEM, Kocaeli, Türkiye}
\IEEEauthorblockA{\IEEEauthorrefmark{2}Department of Electronics and Communication Engineering, Istanbul Technical University, {\.{I}}stanbul, Türkiye}
\IEEEauthorblockA{\IEEEauthorrefmark{3}Department of Electronics and Communications Engineering, Y{{\i}}ld{{\i}}z Technical University, {\.{I}}stanbul, Türkiye\\ 
Emails: \text{\{ubeydullah.erdemir, batuhan.kaplan, ibrahim.hokelek, ali.gorcin\}@tubitak.gov.tr}, \\ \text{agorcin@yildiz.edu.tr, hakan.cirpan@itu.edu.tr}}}
\maketitle

\begin{abstract}
This paper presents field measurement-based channel characterization for air--to--ground (A2G) and air--to--air (A2A) wireless communication systems using two drones equipped with lightweight software-defined radios. A correlation-based channel sounder is employed such that the transmitting drone broadcasts the sounding waveform with a pseudo-noise sequence and the receiving drone captures the sounding waveform together with the location information for the post-processing analysis. The path loss results demonstrate that the measurement and flat-earth two-ray results have similar trends for A2G while the measurement and free space path loss are similar to each other for A2A. The time delays between the direct path and multipath components are widely spread for A2A while the multipath components are mostly concentrated around the direct path for A2G generating a more challenging communication environment. We observe that the reflections from several buildings having metal roofs and claddings on the measurement site cause sudden peaks in the root-mean-square delay spread. The results indicate that the A2A channel has better characteristics than the A2G under similar mobility conditions.
\end{abstract}


\section{Introduction}
The usage of \acp{uav} or drones in the popular term has been growing rapidly for many applications such as cargo transportation, smart agriculture, surveillance, public safety, disaster management, science, and military \cite{shakhatreh2019unmanned, zeng2019accessing, mozaffari2019tutorial}. A high data rate wireless communication is an essential component of these systems since typical applications require the transmission of high resolution images and videos from one drone to another or the \ac{gs}. The integration of terrestrial and non-terrestrial systems including \acp{uav} and satellites is a highly popular topic in 5G and beyond networks. It is an outmost important task to characterize a wideband wireless communication channel for \ac{a2g} and \ac{a2a} operating at 5G carrier frequencies in order to design robust and reliable waveforms for these environments.

A comprehensive survey of the \ac{a2g} channel models for UAVs have been provided in \cite{8740234, 8709739}, where the primary distinction of the \ac{uav} A2G channels is the addition of a third dimension as an altitude. The important parameters include the type of channel sounding signal, its center frequency, bandwidth, transmit power, \ac{uav} speed, heights of \ac{uav} and \ac{gs}, link distance, elevation angle, and local \ac{gs} environment characteristics. The measurement-based models for path loss exponents and shadowing for the radio channel between \acp{uav} and live LTE networks operating at the carrier frequency of 800 MHz are studied in \cite{7936620}. The authors identify that the height of \ac{uav} is an important parameter for characterizing the propagation channel for \acp{uav} as the average number of detected cells increases as the \ac{uav} moves higher altitudes. In \cite{9653074}, the effects of a human body on the \ac{a2g} channel model are studied through measurement experiments when the \ac{uav} is at low altitudes while there are three different use cases of holding a user equipment, namely, near-chest facing, in-pocket facing, and near-chest facing-away. In another recent study \cite{9771480}, the authors characterize \ac{a2g} wireless channels for L-Band (1-2 GHz) and C-Band (4-8 GHz) using a commercially available drone equipped with \ac{sdr}. The drone is used to transmit wideband chirp waveforms whose bandwidths are 26 MHz and 48 MHz for L-band and C-band, respectively, while a commercial signal and spectrum analyzer on the ground is used as the receiver. Our work is similar to the above studies in terms of characterizing \ac{a2g} wireless channels. However, our carrier frequency is 3.5 GHz, the bandwidth is 50 MHz, the drone altitude is 100 meters and we characterize not only \ac{a2g} but also \ac{a2a} wireless channels.

The number of measurement-based channel characterization studies for \ac{a2a} wideband wireless communication channels \cite{9769322, 9135494, 8845191, 9864684} is limited compared to the \ac{a2g} studies. In \cite{9769322}, \ac{a2a} channel measurements are conducted at 5.2 GHz with 100 MHz bandwidth in urban environments by utilizing analog optical links to transmit the channel sounding signal for characterizing the propagation effects, where the exact trajectories of the drones and the three-dimensional layout of the environment are utilized. As an extension study in \cite{9135494}, the authors apply the same measurement campaign for several flight paths in three different environments to find a relation between the multipath components and real-world objects. In \cite{8845191}, field experiments are carried out to measure the effects of the mobility uncertainties on mmWave/THz-band communications between flying drones. Then, the findings are utilized to perform the capacity analysis based on the simulations.

In this paper, field measurement-based channel characterization for \ac{a2g} and \ac{a2a} wireless communication systems is presented using two DJI Matrice 600 Pro drones equipped with lightweight \acp{sdr}. It is a challenging task to perform a measurement-based quantitative characterization of \ac{a2g} and \ac{a2a} wideband wireless communication channels since it requires generating and receiving high-speed data at the transceiver and storing a tremendous amount of data at the receiver. Since the weight of the payload which can be carried by a drone is limited, the \ac{sdr}-based cost-effective and lightweight transceivers are employed for transmitting and receiving the radio signals over the air. A \ac{sbc} with \ac{ssd} is used to generate and store high data rate baseband IQ samples. \ac{gpsdo} is utilized to ensure precise frequency synchronization between the transmitting and receiving \ac{sdr}s. GPS receiver provides high-precision location measurements of both transmitting and receiving drones which are remotely controlled through WLAN interface operating at the carrier frequency of 915 MHz. This control interface is also used to manage the experiments by starting and stopping the measurements. In this study, we employ a correlation-based channel sounder such that the transmitting drone broadcasts the sounding waveform including a \ac{pn}–sequence with the length 4095 and the receiving drone captures the sounding waveform together with the location information for the post-processing analysis. 

The path loss results of the measurement, fit line, \ac{fspl} and \ac{fe2r} models for the \ac{a2g} and \ac{a2a} scenarios demonstrate that the measurement and \ac{fe2r} results have similar trends for A2G while the measurement and \ac{fspl} are similar to each other for \ac{a2g}. The time delay between direct path and multipath components mostly lies between 0 and 300 ns for \ac{a2g} while it is widely spread between 0 and 2 $\mu$s for \ac{a2a}. There are several buildings having metal roofs and claddings on the measurement site, where the reflections from these structures result in sudden peaks in the \ac{rms-ds}. The results indicate that the \ac{a2a} channel has better characteristics than the A2G under similar mobility conditions indicating that higher capacities can be achieved by the \ac{a2a} scenario using the same bandwidth.

\section{Wireless Channel Model}
In this paper, we characterize \ac{a2g} and \ac{a2a} wireless channels through field measurements, where the following model is used for both. A wireless propagation channel including multipath fading is defined as a time-variant system

\begin{equation}
    h(t, \tau) = \sum_{i=0}^{L- 1} a_i(t) \mathrm{e}^{j\phi_i(t)} \delta(\tau - \tau_i(t))
\end{equation}

\noindent where $L$ represents the number of channel taps while $a_i(t)$, $\phi_i(t)$, and $\tau_i(t)$ are the amplitude, phase, and delay of the $i$--th path, respectively. An important measure for a wireless channel is the \ac{pdp}, which characterizes the gains of the respective delays. An instantaneous \ac{pdp} can be calculated as

\begin{equation}
    S(t, \tau) = \left|h\left(t, \tau\right)\right|^2
\end{equation}

\noindent where, $h(t, \tau)$ represents the \ac{cir}. Another important measure that we consider is the path loss model which characterizes the received power with respect to the distance between a transmitter and a receiver. It is expressed as

\begin{equation}
\label{pathloss}
\text{PL}(d)=\underbrace{20\times \log\frac{4\pi d_{0}}{\lambda}}_{\text{PL}_{0}} + 10 \eta \log \left(d / d_{0}\right) + \zeta_{LOS},
\end{equation}

\noindent where $d$ is the distance between a transmitter and a receiver in meters, $\eta$ is the \ac{ple}, $d_{0}$ is the reference distance for path loss measurements, $PL_{0}$ is the \ac{fspl}, and $\zeta_{LOS}$ contains remaining large--scale characteristic losses for \ac{los}. 

Considering the wideband channel modeling, the \ac{rms} delay spread is useful to measure time dispersion or frequency selectivity. The \ac{rms} delay spread can be calculated using the second-order central moment of the \ac{pdp} as

\begin{equation} \label{eq:rms-ds}
    \sigma_\tau = \sqrt{\frac{\sum_{k=0}^{L- 1} a_k^2 \tau_k^2}{\sum_{k=0}^{L- 1} a_k^2} - \left(\frac{\sum_{k=0}^{L- 1} a_k^2 \tau_k^2}{\sum_{k=0}^{L- 1} a_k^2}\right)^2}
\end{equation}

\noindent where $\sigma_\tau$ is the \ac{rms} delay spread for an instantaneous link distance. Also, the latter part of the (\ref{eq:rms-ds}) is defined as the mean excess delay. The reciprocal of the delay spread is the coherence bandwidth and describes the range of frequencies over which two frequency components have a strong potential for amplitude correlation.

\begin{figure}[!b]
    \centering
    \includegraphics[width=\linewidth]{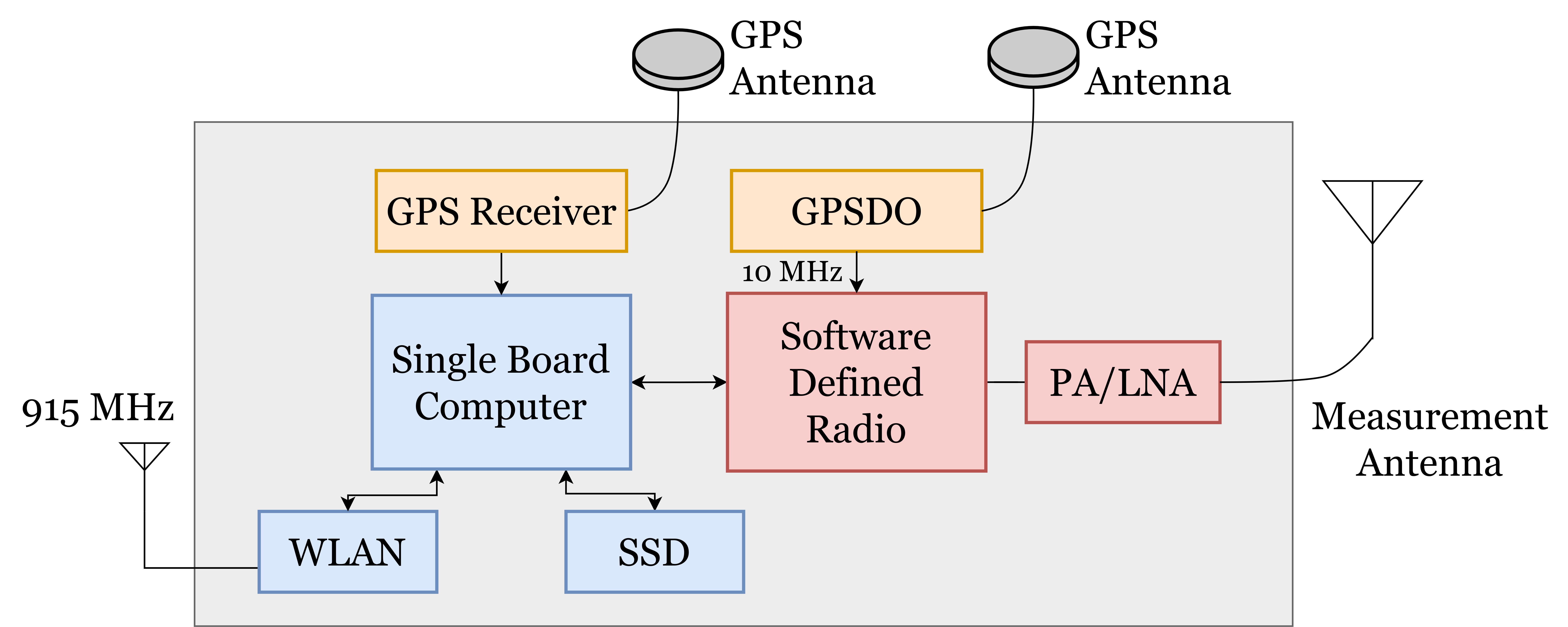}
    \caption{A schematic of the measurement system.}
    \label{fig:schmatic}
\end{figure}

\section{Measurement Campaign}

\subsection{Measurement System}
The channel sounder system model is shown in \FGR{fig:schmatic}, where the transmitter has a power amplifier (PA) whereas the receiver utilizes a low--noise amplifier (LNA). In the transmitter side, first of all, the \ac{sdr} is configured through the \ac{sbc} according to the parameters in \TAB{tab:my-table}. The carrier frequency and the bandwidth are set to 3.5 GHz and 50 MHz, respectively.  The \ac{sbc} continuously generates waveform by reading a predefined \ac{pn} sequence of length 4095 from the \ac{ssd} which is selected to support 1.6 Gbps data transmission need. The \ac{sdr} performs all operations to convert IQ data to RF signals which are amplified up to 30 dBm before transmitting to the antenna.

In the receiver side, before starting the measurement recording, the \ac{sdr} is configured through the \ac{sbc} using the same parameters in the transmitter including the carrier frequency and the bandwidth. During the experiments, the over--the--air signals are collected by the antenna, amplified by the LNA, and sent to the \ac{sdr} which performs all operations to convert the RF signals into IQ data. The \ac{sbc} is configured to hold the 81900 samples of IQ data and writes them periodically to the SSD with a period of 100 ms.

The \ac{gpsdo} provides a 10 MHz clock which ensures precise frequency synchronization between the transmitting and receiving SDRs. The GPS receiver provides high-precision location measurements of both transmitting and receiving drones. The experiment measurements are recorded by adding the locations of measurement points so that the same scenario can be generated for the post-processing analysis. A WLAN interface operating at the carrier frequency of 915 MHz is used to manually and remotely control the channel sounder system.

\begin{figure}
    \centering
    \includegraphics[width=\linewidth]{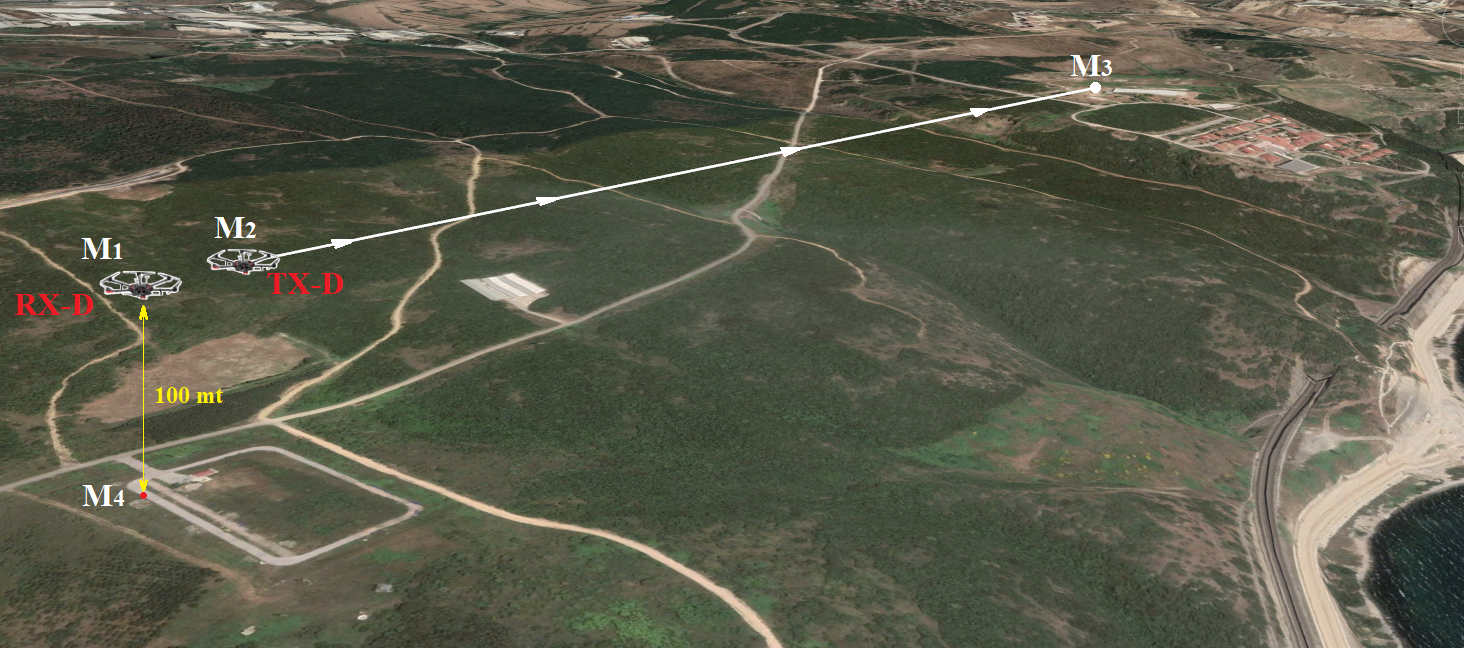}
    \caption{Measurement site and flight route.}
    \label{fig:meas_site}
\end{figure}

\subsection{Measurement Environment}

The measurement experiments are conducted at the TÜBİTAK Gebze campus as shown in \FGR{fig:meas_site}, where two DJI Matrice 600 Pro drones are used. The campus terrain is mostly hilly and forested. The settlement on the campus is concentrated in one area, but there are also sparse building groups in the forested part. The measured environment resembles rural settlements in this sense. The necessary approval for flying drones and performing experiments is obtained from the campus administration. In the first set of experiments, \ac{a2a} measurements are performed such that the \ac{rxd} is located at a measurement point $M_1$ and \ac{txd} is initially located at $M_2$ as depicted in \FGR{fig:meas_site}. Both drones have an altitude of 100 meters and the initial distance between them is 85 meters. The measurements are taken every 100 ms when \ac{rxd} is stationary while \ac{txd} autonomously flies along a 1 km straight trajectory from $M_2$ to $M_3$ in \FGR{fig:meas_site} at a constant speed of 3 m/s at the same altitude.

In the second set of experiments, \ac{a2g} measurements are performed such that the receiver antenna is placed on a mast elongated 3 meters above the ground ($M_4$) while \ac{txd} follows the same trajectory as \ac{a2a} measurement. Note that each experiment takes approximately 6 minutes, which is determined by considering the limited flight time of the drones due to their battery constraints.

\subsection{Measurement Methodology}

\begin{table}[!t]
\centering
\renewcommand{\arraystretch}{1}
\setlength{\tabcolsep}{12pt}
\caption{Waveform Parameters}
\label{tab:my-table}
\resizebox{0.75\columnwidth}{!}{%
\begin{tabular}{l|l}
\toprule
\multicolumn{1}{c}{\textbf{Parameter}} & \multicolumn{1}{c}{\textbf{Value}} \\ \hline
Center Frequency              & $3.5$ GHz                   \\
Bandwidth                     & $50$ MHz                    \\
PN Sequence Length            & $4095$                      \\
Delay Time Resolution         & $20$ ns                     \\
Maximum Delay Time            & $160$ $\mu$s                \\
Transmitted Power             & $30$ dBm                    \\ \hline
\end{tabular}%
}
\end{table}

There are several channel-sounding techniques such as pulse compression, correlation-based, and swept frequency. In this study, a correlation-based channel sounder is employed such that the drone platform needs to carry a low size, weight, and power (SWaP) payload which can only allow transmitting the maximum of 1 Watt RF signal. Therefore, the receiver captures the sounding waveform and the correlation-based channel sounding provides higher processing gain in order to compensate for the low power limitation. The transmitted waveform includes a \ac{pn}--sequence with the length 4095. 
Since a propagation channel is defined as a sum of discrete paths, the correlation function of an $m$--sequence is preferred since it provides higher resolution while distinguishing independent multi-path components \cite{pirkl2008optimal}. 

The measurement procedure is explained as follows. Initially, \ac{txd} starts to broadcast the sounding waveform. After that, \ac{rxd} and \ac{txd} autonomously fly to the measurement points $M_1$ and $M_2$, respectively. We remotely send a start command to \ac{rxd} which begins recording the measurements while \ac{txd} flies through the flight route from $M_2$ to $M_3$. We remotely send a finish command to \ac{rxd} which stops recording the measurements when \ac{txd} arrives $M_3$ and then both drones are landed. During the experiments,  the flight logs of the drones and the IQ records are continuously saved to be used for offline post-processing.

\begin{figure}
    \centering
    \includegraphics[width=\linewidth]{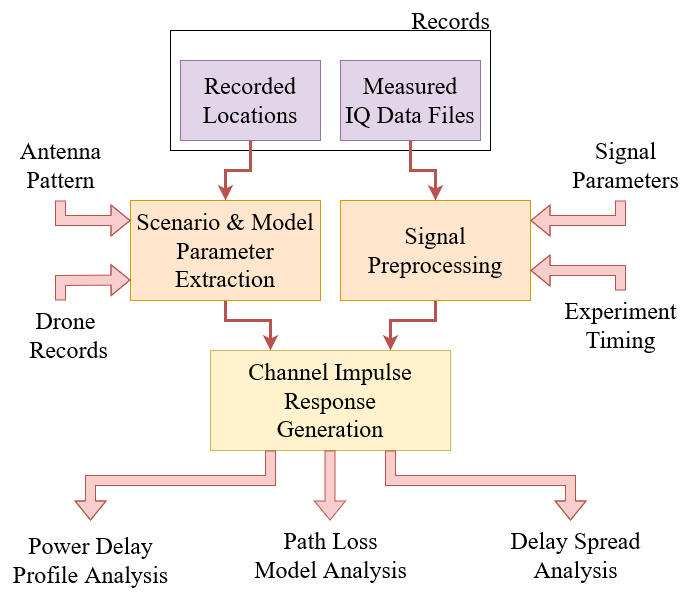}
    \caption{Post-processing Analysis Flow Diagram}
    \label{fig:meth_1}
\end{figure}

The post-processing analysis flow diagram is shown in \FGR{fig:meth_1}. First of all, undesired recordings outside the measurement time interval are removed and the measured IQ data are frequency aligned in the signal pre-processing block. Furthermore, in the scenario and model parameter extraction block, the recorded location information is utilized to obtain the azimuth and elevation angles between TX-D and RX-D. Using the antenna pattern information together with the azimuth and elevation angles, the effects of the antenna patterns are normalized to the omnidirectional pattern. This normalization removes the effects of the antenna on the received signal so that only the channel effects are retained. Then, the losses due to the antenna patterns are calculated and sent to the \ac{cir} generation block which removes these losses on the received signal. 

The received signal includes at least 9 periods of the \ac{pn}--sequence for each measurement which holds the 81900 samples of IQ data. Since the frequency synchronization between TX-D and RX-D is provided by \ac{gpsdo} during the experiments, coherent correlations can be directly performed in the \ac{cir} generation block. A coherent auto-correlation between the raw data and the known transmitted sequence is calculated to extract the \ac{cir} from the raw data. Note that the known transmitted sequence is successively repeated to increase the correlation gain. For example, the correlation gain is 36 dB, 42 dB, and 45 dB when the number of repeated known transmitted sequences is 1, 4, and 8, respectively and it is set to 8 unless otherwise specified. The \ac{cir} contains \ac{dpc}, \ac{mpc}, and noise components. The power threshold, which is used to decide whether an \ac{mpc} is strong or weak, is set to $20$ dB below the \ac{dpc} power \cite{itu_report, Shihao}. And the components that are below the power threshold are excluded from the \ac{cir}. 

\section{Measurement Results \& Discussions}

\begin{figure}[!ht]
\centering
\begin{minipage}[h]{1\linewidth}
    \centering
    \includegraphics[width=1\linewidth]{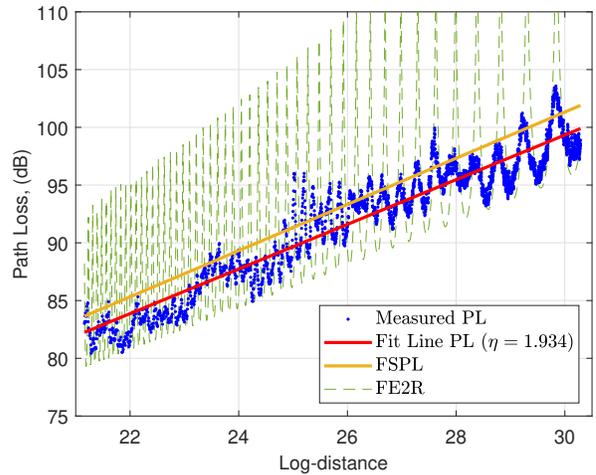}
    \subcaption{A2G}
    \label{Fig_PLa}
\end{minipage} \\
\begin{minipage}[h]{1\linewidth}
    \centering
    \includegraphics[width=1\linewidth]{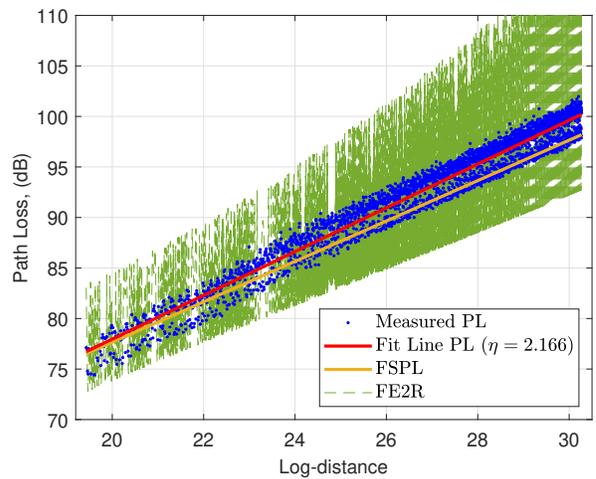}
    \subcaption{A2A}
    \label{Fig_PLb}
\end{minipage}
\caption[]{Path Loss Results.}
\label{fig:PLgroups}
\end{figure}

After the \ac{cir} is generated in \FGR{fig:meth_1}, it can be utilized to obtain important channel characteristics such as path loss model, power delay profile,  and delay spread.
In this section, the measurement results of  these characteristics are presented for \ac{a2a} and \ac{a2g} scenarios.

Fig. \ref{Fig_PLa} shows the path loss results of the measurement, fit line, \ac{fspl} and \ac{fe2r} models for the \ac{a2g} scenario. The path loss measurement values are directly calculated from the \ac{cir} values corresponding to various distances between \ac{txd} and \ac{rxd}. The reference distance is 1 meter.
The fit line results are calculated by fitting the measurement values into the log-distance path loss model defined in \EQ{pathloss}, where the $\eta$ and $\text{PL}_0$ are calculated as 1.934 and 41.320, respectively. The \ac{fe2r} model in \cite{Parsons_2001} is used as a reference since it has been heavily utilized in the literature. The measurement results have a similar trend compared to \ac{fe2r} such that path losses increase with respect to the distance. Since \ac{fe2r} represents an ideal environment where there is only a single reflection which causes significant increases in PL when the phase of the reflected signal is opposite of the phase of the direct-path signal. However, the measurement environment consists of many reflection sources generating different phases and therefore a single phase of a reflection source does not have a significant impact on PL as much as \ac{fe2r}.

Fig. \ref{Fig_PLb} shows the path loss results of the measurement, fit line, \ac{fspl} and \ac{fe2r} models for the \ac{a2a} scenario. For the fit line results, the $\eta$ and $\text{PL}_0$ are calculated as 2.166 and 34.650, respectively. Similar to the \ac{a2g}, the measured PL values for \ac{a2a} increases with respect to distance. The variation of the measured PL values is relatively lower for \ac{a2a} since the scatterers are far from the drones; hence, their effects on the PL values are limited. This observation is also supported by the measured \ac{pdp} results in \FGR{fig:PDPgroups}, where \ac{dpc} is visually distinguishable from the ground reflection and their power strengths are lower for the \ac{a2a}.

\begin{figure}
    \centering
    \includegraphics[width=1\linewidth]{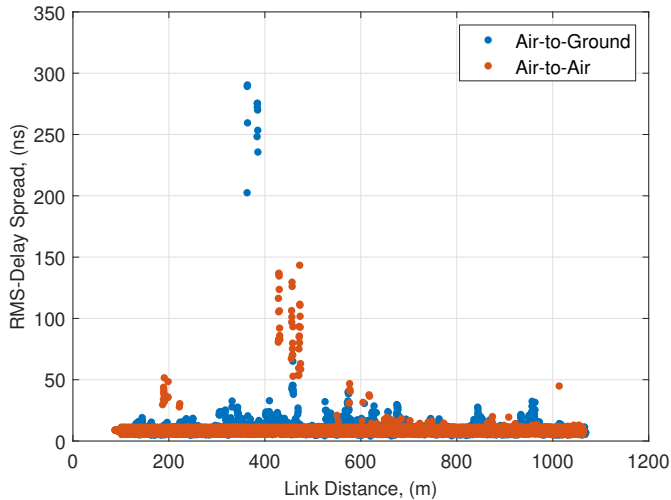}
    \caption{RMS Delay Spread Results.}
    \label{fig:rms-ds}
\end{figure}

\begin{figure}[!t]
\centering
\vspace{-0.3cm}
\begin{minipage}[h]{1\linewidth}
    \centering
    \includegraphics[width=1\linewidth]{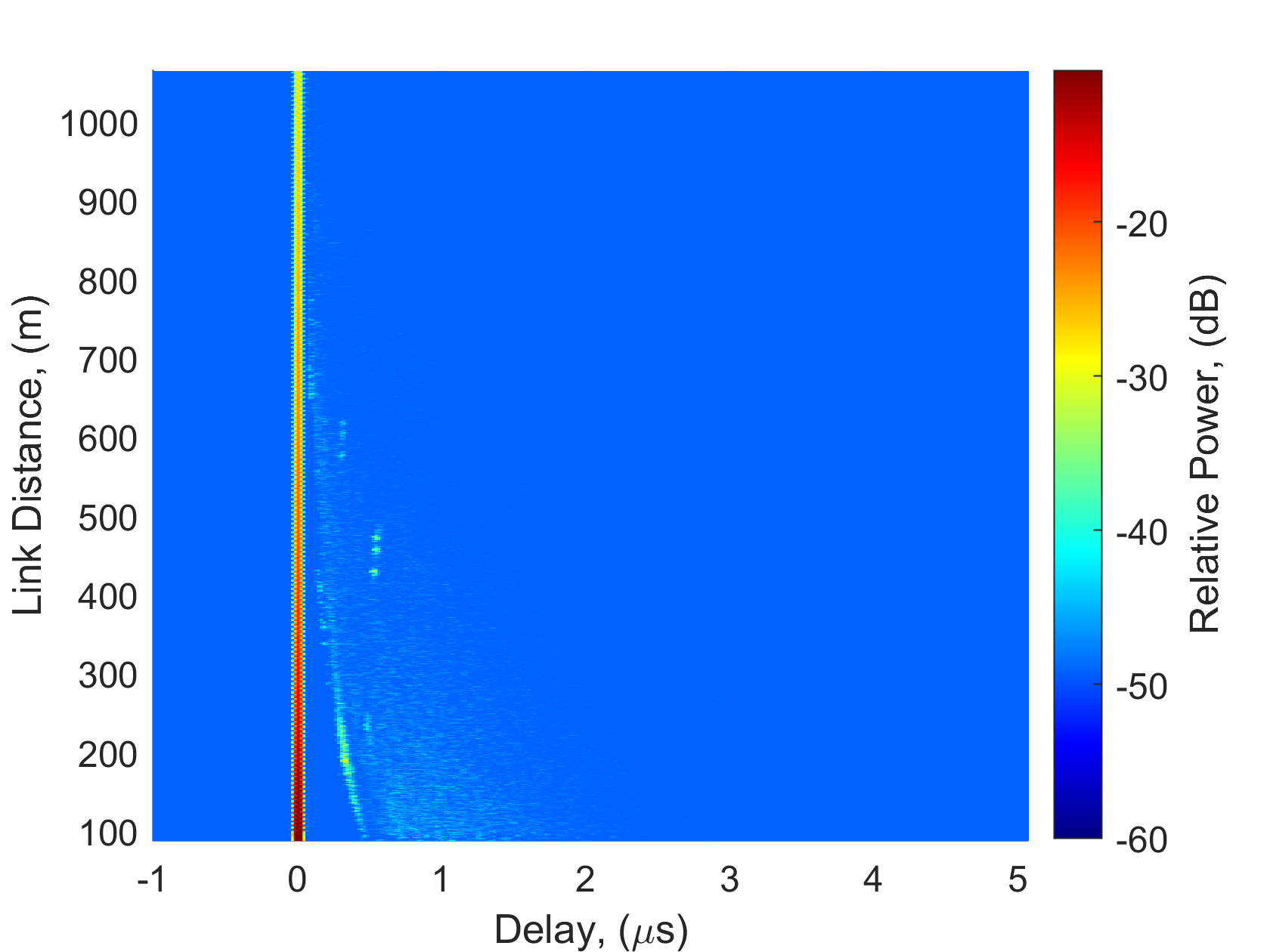}
    \subcaption{A2A PDP.}
    \label{Fig_PDPa}
\end{minipage} \\ 
\begin{minipage}[h]{1\linewidth}
    \centering 
    \includegraphics[width=1\linewidth]{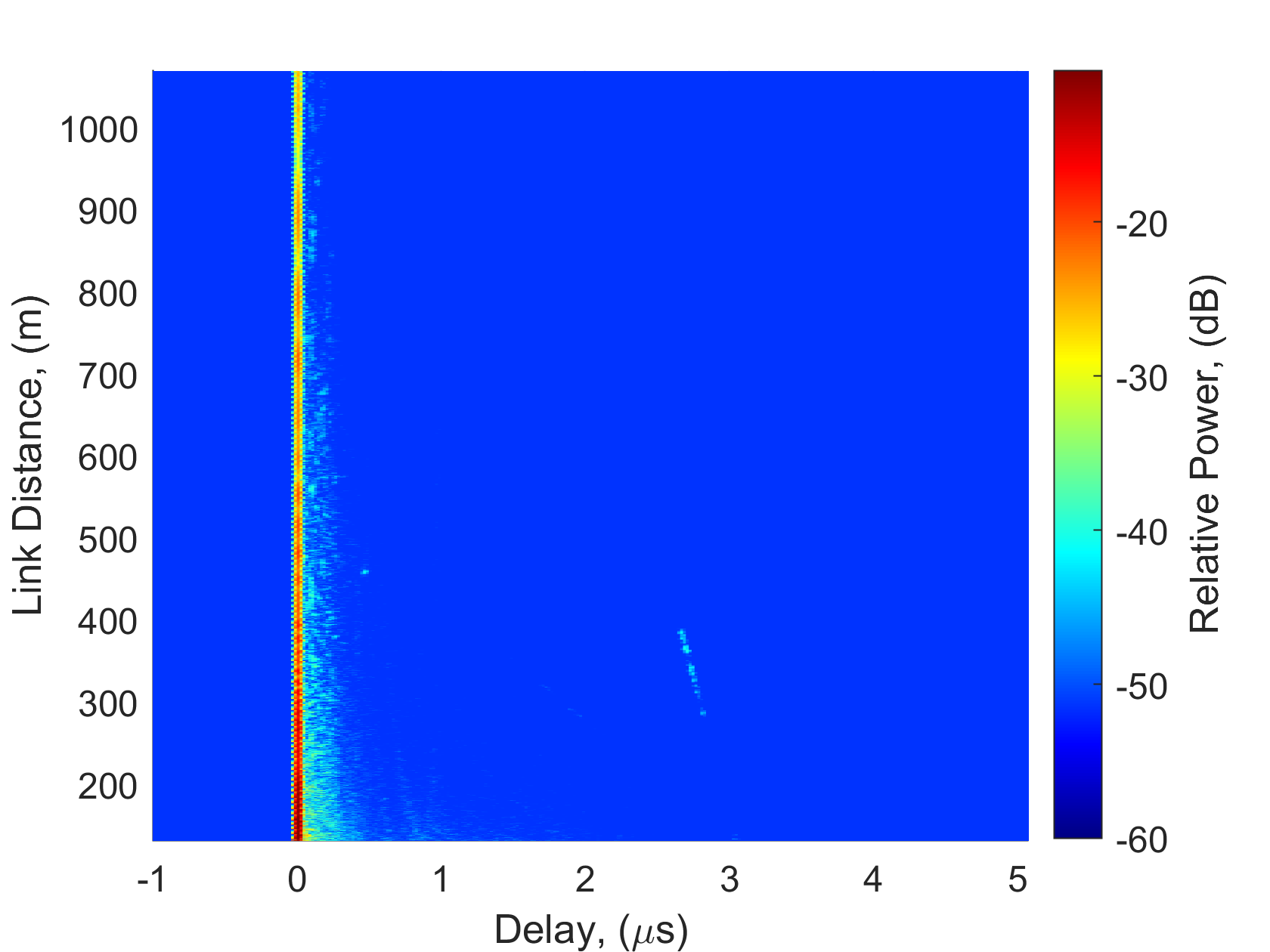}
    \subcaption{A2G PDP.}
    \label{Fig_PDPb}
\end{minipage}
\caption[]{PDP Results.}
\label{fig:PDPgroups}
\end{figure}

\begin{table}
\centering
\renewcommand{\arraystretch}{1.2}
\setlength{\tabcolsep}{14pt}
\caption{Measurement Results of A2G and A2A Scenarios for Drone Channel Characterization}
\label{tab:allResult}
\resizebox{\columnwidth}{!}{%
\begin{tabular}{lccc}
\toprule
\multicolumn{1}{l}{} & \multicolumn{1}{l}{} & \textbf{A2G} & \textbf{A2A} \\ \hline
\multicolumn{1}{l|}{$\eta$} & \multicolumn{1}{l}{} & $1.934$ & $2.166$ \\
\multicolumn{1}{l|}{$\text{PL}_0$ (dB)} & \multicolumn{1}{l}{} & $41.320$ & $34.650$ \\ \hline
\multicolumn{1}{l|}{\multirow{5}{*}{$\sigma_\tau $ (ns)}} & Mean & 10.983 & 9.884 \\ \cline{2-4} 
\multicolumn{1}{c|}{} & \begin{tabular}[c]{@{}c@{}}Standart\\ Deviation\end{tabular} & $15.243$ & $10.051$ \\ \cline{2-4} 
\multicolumn{1}{c|}{} & Median & $9.369$ & $8.676$ \\ \cline{2-4} 
\multicolumn{1}{c|}{} & Min & $4.073$ & $4.806$ \\ \cline{2-4} 
\multicolumn{1}{c|}{} & Max & $290.389$ & $143.405$ \\ \hline
\end{tabular}%
}
\end{table}

\FGR{fig:rms-ds} shows \ac{rms-ds} ($\sigma_\tau$) results for both \ac{a2g} and \ac{a2a}, where $\sigma_\tau$ values are mostly lower than 20 ns and 50 ns for \ac{a2a} and \ac{a2g}, respectively. However, sudden peaks are observed for  $\sigma_\tau$ when the distance is between 300-400 m and 400-500 m for \ac{a2g} and \ac{a2a}, respectively. These peaks are due to either a strong reflection from a near source or a weak reflection from a distant source as predicted by \EQ{eq:rms-ds}. There are several buildings having metal roofs and claddings on the measurement site. These structures cause the reflections resulting in sudden peaks in the \ac{rms-ds}. Note that $\sigma_\tau$ can be up to 144 ns for \ac{a2a} while it is up to 291 ns for \ac{a2g}. The statistics of $\sigma_\tau$ are summarized in \TAB{tab:allResult}.

The measured PDPs for \ac{a2a} and \ac{a2g} are shown in \FGRu{Fig_PDPa} and \ref{Fig_PDPb}, respectively. For the \ac{a2a} scenario, the \ac{mpc} due to the main ground reflection is visually distinguishable when the distance between the drones is close. As \ac{txd} moves away, the time delay between the \ac{mpc} and the \ac{dpc} decreases as expected. The other reflections from the forested terrain smoothly fade as depicted in \FGR{Fig_PDPa}. The dominant reflections in the figure are likely caused by the buildings having metal roofs and claddings as explained above. The time delay between \ac{dpc} and \acp{mpc} mostly lies between 0 and 300 ns for \ac{a2g} while it is widely spread between 0 and 2 $\mu$s for \ac{a2a}. However, the power of the \acp{mpc} is significantly higher for \ac{a2g} resulting in higher \ac{rms-ds}. The time delay of the \acp{mpc} causing the sudden peaks in \ac{rms-ds} as depicted in \FGR{fig:rms-ds} can be visually seen in \FGR{fig:PDPgroups}.

\section{Conclusion}
In this paper, we presented field measurement-based channel characterization for \ac{a2g} and \ac{a2a} wireless communication systems using two drones hosting lightweight \ac{sdr}. A sounding waveform with a \ac{pn} sequence is sent by the transmitting drone and the receiving drone captures the sounding waveform to extract several channel characteristics such as path loss, \ac{rms-ds}, and multipath delays. The path loss results for the measurement and \ac{fe2r} show similar trends for \ac{a2g} while the measurement and \ac{fspl} are similar for \ac{a2a}. The multipath delays create more challenging channel conditions for \ac{a2g} under a similar mobility scenario.
Extensive measurements will be made with the sounder in different environments and waveform parameters for future work.

\section*{Acknowledgment}
We thank to StorAIge project that has received funding from the KDT Joint Undertaking (JU) under Grant Agreement No. 101007321. The JU receives support from the European Union’s Horizon 2020 research and innovation programme in France, Belgium, Czech Republic, Germany, Italy, Sweden, Switzerland, Türkiye, and National Authority TÜBİTAK with project ID 121N350.

\balance
\bibliographystyle{IEEEtran}
\bibliography{references.bib}

\end{document}